\documentclass[superscriptaddress,showpacs,preprintnumbers,amsmath,amssymb,twocolumn,prb]{revtex4-1}
\usepackage{graphicx}
\usepackage{dcolumn}
\usepackage{bm}
\usepackage{color}
\usepackage{epstopdf}
\renewcommand{\bm}{\mathbf}
\begin{document}

\title{Exciton-polaritons in CsPbBr$_3$ crystals revealed by optical reflectivity in high magnetic fields and two-photon spectroscopy}
% \\ Exciton parameters in CsPbBr$_3$ crystal revealed by high magnetic fields and two-photon spectroscopy}  
%\\ Exciton-polaritons in CsPbBr$_3$ revealed by high magnetic fields and two-photon spectroscopy}

\author{Dmitri R. Yakovlev} 
\email{dmitri.yakovlev@tu-dortmund.de}
\affiliation{Experimentelle Physik 2, Technische Universit{\"a}t Dortmund, 44227 Dortmund, Germany}
%\affiliation{Ioffe Institute, Russian Academy of Sciences, 194021 St. Petersburg, Russia}

\author{Scott A. Crooker}
%\email{crooker@lanl.gov}
\affiliation{National High Magnetic Field Laboratory, Los Alamos National Lab, Los Alamos, New Mexico 87545, USA}

\author{Marina A. Semina}
\affiliation{Ioffe  Institute, Russian Academy of Sciences, 194021 St. Petersburg, Russia}
%\email{msemina@gmail.com}

\author{Janina Rautert}
\affiliation{Experimentelle Physik 2, Technische Universit{\"a}t Dortmund, 44227 Dortmund, Germany}

\author{Johannes Mund}
\affiliation{Experimentelle Physik 2, Technische Universit{\"a}t Dortmund, 44227 Dortmund, Germany}

\author{Dmitry N. Dirin}
\affiliation{Laboratory of Inorganic Chemistry, Department of Chemistry and Applied Biosciences, ETH Z{\"u}rich, CH-8093 Z{\"u}rich, Switzerland}
\affiliation{Laboratory for Thin Films and Photovoltaics, Department of Advanced Materials and Surfaces, Empa - Swiss Federal Laboratories for Materials Science and Technology, CH-8600 D{\"u}bendorf, Switzerland}

\author{Maksym V. Kovalenko}
\affiliation{Laboratory of Inorganic Chemistry, Department of Chemistry and Applied Biosciences, ETH Z{\"u}rich, CH-8093 Z{\"u}rich, Switzerland}
\affiliation{Laboratory for Thin Films and Photovoltaics, Department of Advanced Materials and Surfaces, Empa - Swiss Federal Laboratories for Materials Science and Technology, CH-8600 D{\"u}bendorf, Switzerland}

\author{Manfred Bayer}
\affiliation{Experimentelle Physik 2, Technische Universit{\"a}t Dortmund, 44227 Dortmund, Germany}
%\affiliation{Ioffe  Institute, Russian Academy of Sciences, 194021 St. Petersburg, Russia}

\date{\today}

\begin{abstract}
Cesium lead bromide (CsPbBr$_3$) is a representative material of the emerging class of lead halide perovskite semiconductors that possess remarkable optoelectronic properties. Its optical properties in the vicinity of the band gap energy are greatly contributed by excitons, which form exciton-polaritons due to strong light-matter interactions. We examine exciton-polaritons in solution-grown CsPbBr$_3$ crystals by means of circularly-polarized reflection spectroscopy measured in high magnetic fields up to 60~T. The excited 2P exciton state is measured by two-photon absorption. Comprehensive modeling and analysis provides detailed quantitative information about the exciton-polariton parameters: exciton binding energy of 32.5~meV, oscillator strength characterized by longitudinal-tranverse splitting of 5.3~meV, damping of 6.7~meV, reduced exciton mass of $0.18 m_0$, exciton diamagnetic shift of 1.6~$\mu$eV/T$^2$, and exciton Land\'e factor $g_X=+2.35$. We show that the exciton states can be well described within a hydrogen-like model with an effective dielectric constant of 8.7. From the measured exciton longitudinal-transverse splitting we evaluate the Kane energy of $E_p=15$~eV, which is in reasonable agreement with values of $11.8-12.5$~eV derived from the carrier effective masses. 

\end{abstract}

\maketitle


%\cD{Paper is planned for ACS Photonics.}

%\textbf{Introduction}\\

Cesium lead bromide (CsPbBr$_3$) is a fully-inorganic member of the large family of lead halide perovskite semiconductors~\cite{Vardeny2022_book,Vinattieri2021_book}. This material delivers an outstanding combination of optoelectronic properties, such as long charge carrier diffusion lengths up to 1.3~$\mu$m,~\cite{Fan2020} high carrier mobilities ($10-181$~cm$^2$ V$^{-1}$s$^{-1}$),~\cite{Fan2020,Zhang2021} low densities of carrier trap states ($3.9\times 10^{10}$~cm$^{-3}$) despite a large density of point defects~\cite{Fan2020}, and small carrier effective masses  ($0.17m_0-0.26m_0$) combined with large polarons~\cite{Puppin2020}. CsPbBr$_3$ has been recently applied to the fabrication of perovskite solar cells~\cite{Kulbak2015,Kulbak2016}, sensitive visible light detectors~\cite{Song2016}, and high-energy detectors~\cite{Stoumpos2013,Matt2013}. Fully inorganic CsPbBr$_3$ exhibits higher stability against elevated temperatures and polar solvents compared to analogous lead halide perovskites that incorporate organic cations, e.g., methyl-ammonium (MA) lead bromide. In applications for hard radiation detection, CsPbBr$_3$ is indispensable as it is the only lead halide perovskite that can operate under the high bias required for efficient extraction of charge carriers~\cite{Stoumpos2013}.
 
Optical properties of semiconductors in the vicinity of the band gap energy are controlled by excitons, making  knowledge of the exciton parameters of great importance for optoelectronics. CsPbBr$_3$  is a model material for the class of lead halide perovskite semiconductors. It is widely studied experimentally and theoretically, with initial papers on exciton properties dating back to 1978~\cite{Heidrich78,Ito78,Froelich1979,Heidrich81}. Light-matter interactions are quite strong in CsPbBr$_3$~\cite{Su2020,Su2021,Tao2022,Dursun2018}, which results in a pronounced resonance in reflectivity and absorption spectra in the vicinity of the band gap, with features of exciton-polaritons~\cite{Klingshirn_book_2005}. This interaction can be further enhanced in a microcavity~\cite{Bao2019}. 

Magneto-optical methods are powerful tools to measure exciton parameters in semiconductors~\cite{Ivchenko_book_2007, Seisyan_1991, Seisyan_2012, Reynolds_book_1981}. Exciton binding energies can be evaluated from the diamagnetic shift of the 1S exciton state in magnetic fields and/or from the observation of excited exciton states (2S, 3S, etc.), which gain oscillator strength and become increasingly visible in high fields. In both cases the exciton linewidth is a critical parameter for the evaluation accuracy, which in turn is improved in high magnetic fields. Magneto-optical studies in high magnetic fields of $45-150$~T were successfully used for investigation of  lead halide perovskite semiconductors~\cite{Kataoka1993, Hirasawa1994, Tanaka2003,Tanaka2005,Miyata2015, Galkowski2016, Yang2017,Yang2017b,Baranowski2019,Baranowski2020a, Baranowski2020AEM}. Most studies to date were made on polycrystalline films with rather broad exciton lines, and results on single crystals with small inhomogeneous broadening of the excitons are available only for MAPbBr$_3$~\cite{Baranowski2019} and MAPbI$_3$~\cite{Yang2017b}.

For CsPbBr$_3$ films, measurements in magnetic fields up to 150~T indicate an exciton binding (Rydberg) energy of 33~meV and a reduced exciton mass of $\mu=0.126 m_0$~\cite{Yang2017}. Note that earlier and not very detailed reports inferred exciton binding energies of 37~meV~\cite{Froelich1979} and 34~meV~\cite{Pashuk1981}.   Importantly, however, CsPbBr$_3$ has a rich spectrum of optical phonons with energies ranging from 4 to 25~meV~\cite{Guo2019,Zhou2020}, values which are close to, but smaller than, the exciton binding energy. It is generally in the energy range of the optical phonons where the magnitude of the energy-dependent dielectric constant $\varepsilon(\omega)$ varies significantly between its stationary (i.e. zero-frequency) value ($\varepsilon_s$) and its value in the high-frequency limit ($\varepsilon_\infty$). This fact provides two aspects for the exciton properties and calculation of its parameters. The first one is that phenomenologically the exciton energies are often described in the framework of a hydrogen-like model, but with an effective dielectric constant $\varepsilon_{eff}$, which falls between the high frequency and static dielectric constants $\varepsilon_{\infty}<\varepsilon_{eff}<\varepsilon_{s}$. The second one is that the dielectric screening of the exciton may, in fact, differ between the exciton ground and excited states due to their different radii~\cite{Zunger} and, consequently, binding energies, such that the energy spectrum of excited exciton states may not conform to the classic Rydberg series that is predicted by hydrogen-like models (which assume a constant $\varepsilon_{eff}$). Note, that the exciton spectra also can differ from the Rydberg series in semiconductors with complex band structure~\cite{Lipari71,Lipari73,Lipari74}.  Although, in the lead halide perovskites the top valence and bottom conduction bands have a simple structure~\cite{Even2012, kirstein2021nc} (see also Figure~S1a),  making the hydrogenic model a good approximation at first glance.  

%, for details see Supporting Information. \cS{does the supporting information actually say anything about how the 'complex band structure can lead to an anomalous Rydberg series? I don't see that the SI addresses this point.   If not, maybe this sentence and the following one are distracting and unnecessary (?).} 

In this study we check the applicability of the Rydberg formula for excitons in CsPbBr$_3$ and determine fundamental exciton parameters such as the Rydberg energy, reduced mass, and effective dielectric constant. The reflectivity of light in the optical range of the exciton ground state in high magnetic fields up to 60~T is measured for two circular polarizations of incident light. We model the measured spectra within the exciton-polariton model allowing us to determine with a high accuracy the positions of transverse and longitudinal excitons and their energy splitting, $g$-factors, and diamagnetic shift. To make the evaluation of exciton parameters more accurate the excited 2P exciton state has also been measured by two-photon excitation in magnetic fields up to 10~T.

%%%%%%%%%%%%%%%%%%%%%%%%%%%%%%%%%%%%%%%%%%%%%%%%%%%%%%%%%%%%%%%%%%%%%%%%%%%%%%%%%%%%%%

\section*{Results}
\label{sec:experimental_results}

The sample under study is a  fully inorganic lead halide perovskite  bulk crystal of CsPbBr$_3$ grown in solution (Methods). Information on its optical and spin properties in low magnetic fields can be found in Refs.~\cite{Belykh2019, kirstein2021nc, kopteva2023_gX}.  Here we focus on exciton-polariton parameters and their modification in strong magnetic fields. Experiments were performed at $T=1.6$~K, with the sample immersed in pumped liquid helium.

It has been appreciated for decades~\cite{Hopfield,Andreani1988, Klingshirn_book_2005,Reynolds_book_1981} that a proper description of excitons interacting with light in semiconductors requires consideration of the exciton-polariton effect, which accounts for the interaction of the mechanical exciton with the light wave. This effect characterizes the dependence of the dielectric function on the wave vector (spatial dispersion) and gives rise to two exciton-polariton modes, which modifies the optical properties  of the semiconductor in the spectral range of the exciton-polariton resonance. The splitting between transverse and longitudinal (with respect to the wave vector)  excitons is the consequence of the long-range part of the exchange interaction \cite{Andreani1988} and, simultaneously, is connected with the exciton polarizability, describing its oscillator strength and playing the key role in exciton-polariton dispersion relation.  Here we analyze the optical reflectivity of CsPbBr$_3$ crystals in the context of the exciton-polariton model and derive fundamental optoelectronic parameters.

\textbf{Reflectivity of exciton-polaritons in pulsed magnetic fields.} 

\begin{figure}[hbt]
\includegraphics[width=0.97\columnwidth]{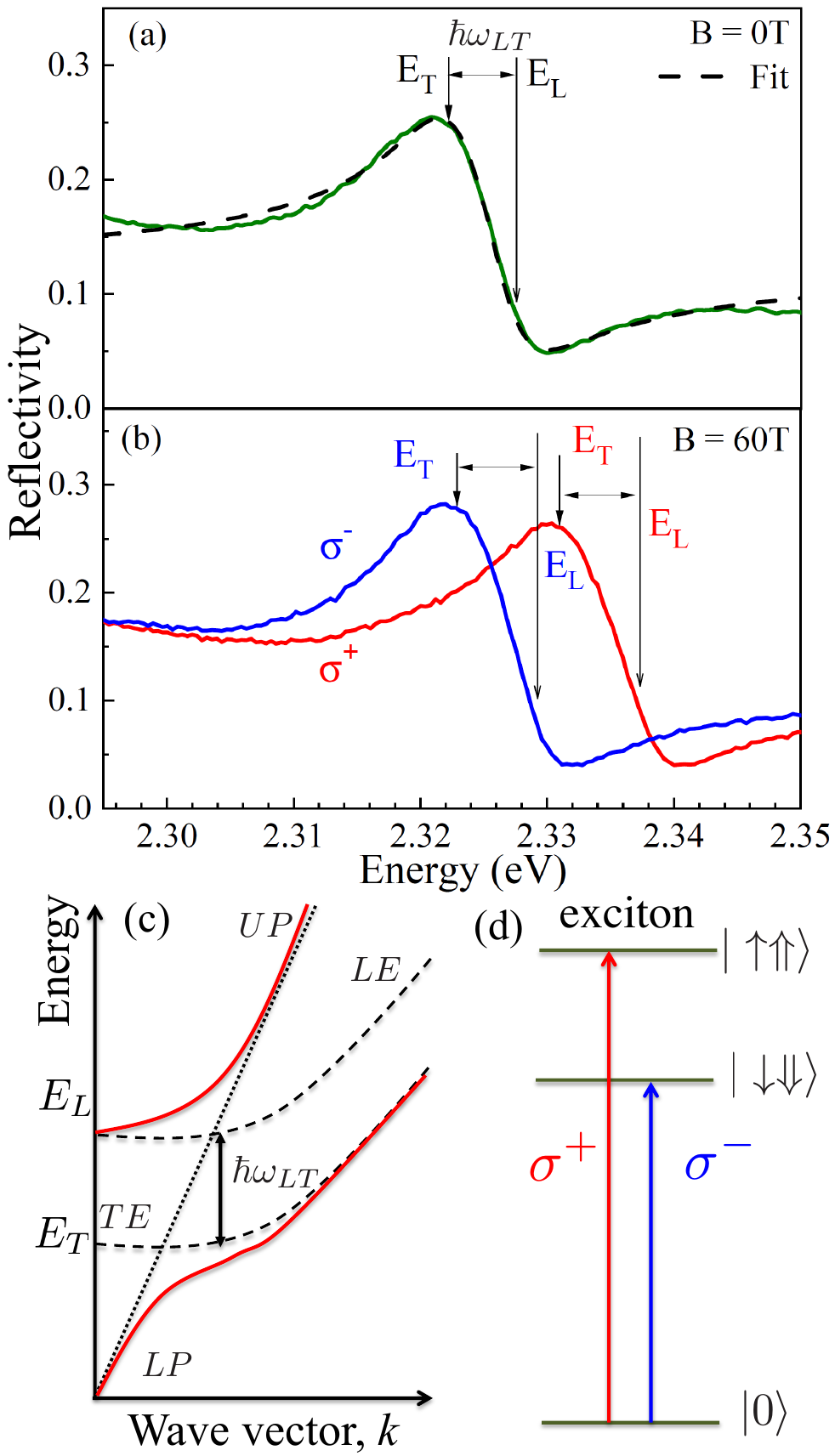}
   \caption{Reflectivity spectra of CsPbBr$_3$ in (a) zero and (b) 60~T magnetic field, measured at $T=1.6$~K. Solid lines are experiment and the dashed line is a fit using $E_T=2.322$~eV, $\hbar\omega_{LT}=5.3$~meV, $\hbar\Gamma=6.7$~meV,  and $L=2.3$~nm.  (c) Scheme of the exciton-polariton dispersion that accounts for the spatial dispersion.  (d)  Scheme of the exciton optical transitions in magnetic field. Single and double arrows correspond to electron and hole spins, respectively, which in optically active exciton states have  spin $+1$ for $|\uparrow \Uparrow  \rangle$  and $-1$ for $|\downarrow \Downarrow  \rangle$.  $|0 \rangle$ is the ground state of the unexcited crystal. }
 \label{fig:1}
\end{figure}

The reflectivity spectrum of the CsPbBr$_3$ crystal at zero magnetic field is shown in Figure~\ref{fig:1}a. It has a  pronounced exciton-polariton resonance, which results from the interaction of the mechanical excitons with the photons and their conversion to each other in the crystal. For its model description and fit of the reflectivity spectrum we use   the commonly accepted theory of exciton-polaritons~\cite{Hopfield, Klingshirn_book_2005}, see Supporting Information S2. For the boundary conditions we take the ``dead layer'' approach introduced by Hopfield and Thomas~\cite{Hopfield}. Schematically the exciton-polariton dispersion is shown in Fig.~\ref{fig:1}c. Here, $E_T$ is the energy of the transverse (``mechanical'') exciton, $E_L$ is the energy of the longitudinal exciton, and $\hbar \omega_{LT}=E_{L}-E_T$ is the longitudinal-transverse splitting, which characterizes the exciton oscillator strength \cite{Denisov}. In Fig.~\ref{fig:1}c ``LP'' and ``UP'' denote lower and upper exciton-polariton branches, dashed line ``LE'' corresponds to longitudinal exciton, dashed line ``TE'' to transverse exciton without account of the polariton effect, dotted line shows the light dispersion. In cubic crystals longitudinal excitons are optically inactive and at zero wave vector their energy coincides with the upper polariton branch~\cite{Hopfield60,Andreani1988}. Although,  CsPbBr$_3$  in low temperatures is in the orthorombic phase, which has a lower symmetry than cubic, and, in principle, longitudinal exciton may be observable in optics, in experiment we do not see any line that can be attributed to LE. 

 The dashed line in Fig.~\ref{fig:1}a shows the fit of the experimental reflectivity spectrum with Eq.~(S4).  Here we took  the background dielectric constant $\varepsilon_b=4.3$\cite{Miyata2017} as a high frequency one  and the translation exciton mass $M=0.72 m_0$. Note, that the fit does not depend strongly on the value of $M$, for details see Supporting Information S2.   The best fit provides the following exciton-polariton parameters: $E_T=2.322$~eV, $\hbar\omega_{LT}=5.3$~meV, exciton damping $\hbar\Gamma=6.7$~meV, and ``dead'' layer thickness $L=2.3$~nm. We note the considerable value of $\hbar\omega_{LT}$, which justifies the account of the polariton effect in modeling the reflectivity spectra of CsPbBr$_3$. 

Reflectivity spectra were measured in pulsed magnetic fields up to $60$~T in $\sigma^+$ and $\sigma^-$ circular polarizations in order to distinguish exciton states with spin $\pm1$ and evaluate the exciton Zeeman splitting and its Land\'e $g$-factor (see scheme in Fig.~\ref{fig:1}d). The spectra measured in $B=60$~T are shown in Fig.~\ref{fig:1}b. One can see the pronounced Zeeman splitting of the oppositely polarized exciton resonances. 

\begin{figure*}[hbt]
\begin{center}
\includegraphics[width = 16cm]{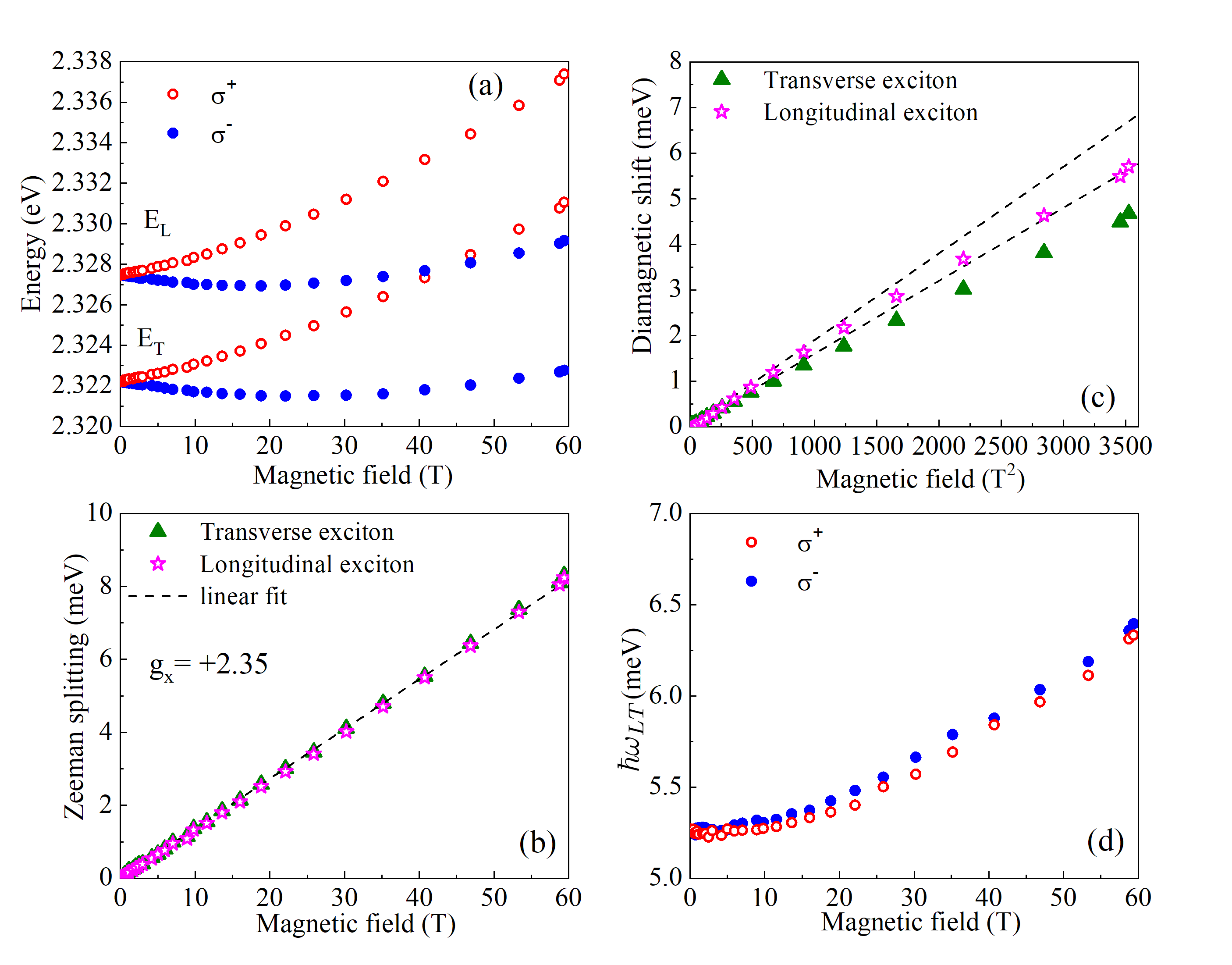}
 \caption{Parameters of the 1S exciton-polariton in CsPbBr$_3$ in strong magnetic fields evaluated from polarized reflectivity spectra.  $T=1.6$~K. (a) Energies of transverse $(E_T)$ and longitudinal $(E_L)$ excitons active in $\sigma^+$ and $\sigma^-$ circular polarization as a function of magnetic field. (b) Magnetic field dependence of the Zeeman splitting $E_{\sigma^+ }-E_{\sigma^-}$ of transverse (triangles) and longitudinal (open stars) excitons. The dashed line is a linear fit with $g_{X}=+2.35$.  (c)  Diamagnetic shift  of transverse and longitudinal excitons vs. $B^2$. Dashed lines are $B^2$ fits with $c_d^{1S}=1.6$~$\mu$eV/T$^2$ and $c_d^{L}=2$~$\mu$eV/T$^2$, respectively. (d) The longitudinal-transverse splitting in two circular polarizations as a function of magnetic field.}
\label{fig:2}
\end{center}
\end{figure*}

There are three effects of magnetic fields on the exciton resonance that we measured and analyzed: the diamagnetic energy shift (which is same for both spin components), the Zeeman splitting, and the increase of the exciton oscillator strength. In the regime of weak magnetic fields, where the exciton binding energy exceeds the cyclotron energies of the electrons and holes, the energy of the $n$-th exciton state in the hydrogen-like model is 
\begin{eqnarray}\label{gamma}
E_X^{(n)} = E_g - \frac{R^*}{n^2} \pm \frac{1}{2} g_X \mu_B B + c_d^{(n)} B^2 \,. 
\end{eqnarray}
Here $E_g$ is the free-particle band gap energy, $g_X$ the exciton Land\'e $g$-factor, and $\mu_B$ is the Bohr magneton. Here the plus sign in the Zeeman term $\pm \frac{1}{2} g_X \mu_B B $ corresponds to excitons optically active in $\sigma^+$ polarization, and the minus sign corresponds to excitons active in $\sigma^-$. $R^*$ is the exciton Rydberg, which is equal to the binding energy of a 1S hydrogen-like exciton ground state $E_b^{1S}$: 
\begin{eqnarray}\label{Rydberg}
 R^*=E_b^{1S}=\frac{\mu e^4}{2\hbar^2\varepsilon_{{\rm eff}}^2} = \frac{e^2}{2\varepsilon_{{\rm eff}} a_B} \,, 
\end{eqnarray}
where $\mu$ is the reduced exciton mass, which is composed of the electron and hole effective masses as $\mu^{-1}=m_e^{-1}+m_h^{-1}$, $e$  is the electron charge, $\hbar$ is the Planck constant, $\varepsilon_{{\rm eff}}$ is the effective dielectric constant, and $a_B$ is the exciton Bohr radius:  
\begin{eqnarray}\label{Bohr_radius}
 a_B=\frac{\hbar^2 \varepsilon_{\rm eff}}{\mu e^2 } \,. 
\end{eqnarray}
 The diamagnetic coefficient $c_d^{(n)}$ is given by
\begin{eqnarray}\label{gamma2}
 c_d^{(n)}= \frac{e^2}{8\mu c^2} \langle\rho_n^2 \rangle \,, 
\end{eqnarray}
where $c$ is speed of light and $\langle\rho_n^2 \rangle$ is the mean square of exciton wave function size in the plane perpendicular to the magnetic field. In the hydrogen-like model for the 1S-state, $\langle\rho_{1S}^2 \rangle = 2a^2_B$ and for the 2P$_{10}$-state $\langle\rho_{2P}^2 \rangle = 12a^2_B$.
% \cD{Marina, what means low index "10" here? Do we need it? Seems that we use it only here.} 
The binding energy of the 2P state is 4 times smaller than the binding energy of the 1S state:
\begin{equation}\label{EBP2}
E_b^{\rm 2P}=\frac{\mu e^4}{8\hbar^2\varepsilon_{{\rm eff}}^2}=\frac{e^2}{8\varepsilon_{{\rm eff}} a_B}.
\end{equation}

We fit the reflectivity spectra measured at all magnetic fields and plot the field dependence of these parameters in Fig.~\ref{fig:2}. Figure~\ref{fig:2}a shows the $E_T$ and $E_L$ energies of the exciton states active in $\sigma^+$ and $\sigma^-$ circular polarizations. The energy  of $E_L$ we calculate as $E_L=E_T+\hbar\omega_{LT}$. The Zeeman splitting of the exciton states calculated as $E_{\sigma^+ }-E_{\sigma^-}= g_X \mu_B B$ is shown in Fig.~\ref{fig:2}b. The dependences coincide for the $E_T$ and $E_L$ excitons,  and their linear fit gives the exciton $g$-factor $g_{X}=+2.35$. We emphasize that it is a direct measurement of $g_X$ value and sign, which is positive. This result is in agreement with our recent reports on the universal dependence of exciton~\cite{kopteva2023_gX} and isolated charge carrier~\cite{kirstein2021nc} $g$-factors on the band gap energy in lead halide perovskite bulk crystals. Note that the high linearity of the field dependence of the Zeeman splitting  indicates that band mixing is negligibly small in lead halide perovskites even in very strong magnetic fields. This is explained by the simple spin structure (spin 1/2) of the electron and hole bands in the vicinity of the band gap (which contribute primarily to the exciton wave function), and the rather large energy separation to the higher-lying conduction bands. 

The diamagnetic shift of the transverse and longitudinal excitons, which was calculated as the center-of-gravity of the $\sigma^\pm$ components, is shown in Fig.~\ref{fig:2}c as a function of the square of magnetic field. Dashed lines  are $B^2$ fits; they  describe well the experimental data in fields up to 25~T (more details will be given below in Fig.~\ref{fig:4}). From the fits we evaluate the diamagnetic coefficients for the transverse exciton $c_d^{1S}=1.6$~$\mu$eV/T$^2$ and longitudinal exciton $c_d^{L}=2$~$\mu$eV/T$^2$. Their difference originates from the increase of the exciton oscillator strength ($\hbar\omega_{LT}$) with magnetic field, which is shown in Fig.~\ref{fig:2}d. It increases from 5.3~meV at zero field up to 6.4~meV at $B=60$~T. The exciton damping $\hbar\Gamma$ does not significantly depend on the magnetic field (not shown here). 

\textbf{Two-photon spectroscopy of 2P exciton state.}

Information about excited exciton states enriches the knowledge of the exciton parameters and allows one to define them with higher precision. Unfortunately, the 2S exciton state was not detectable in reflectivity spectra even in strong magnetic fields, most probably due to its small oscillator strength. Therefore, we used another approach and measured the 2P exciton state which is forbidden for one-photon excitation, but is allowed for two-photon excitation. We measured the photoluminescence excitation (PLE) spectrum using two-photon excitation, i.e. the laser photon energy was tuned in the range close to $E_g/2$. Recently, we used this technique to study excitons in CdSe colloidal nanoplatelets~\cite{Shornikova2021}.

\begin{figure}[hbt]
\includegraphics[width=\columnwidth]{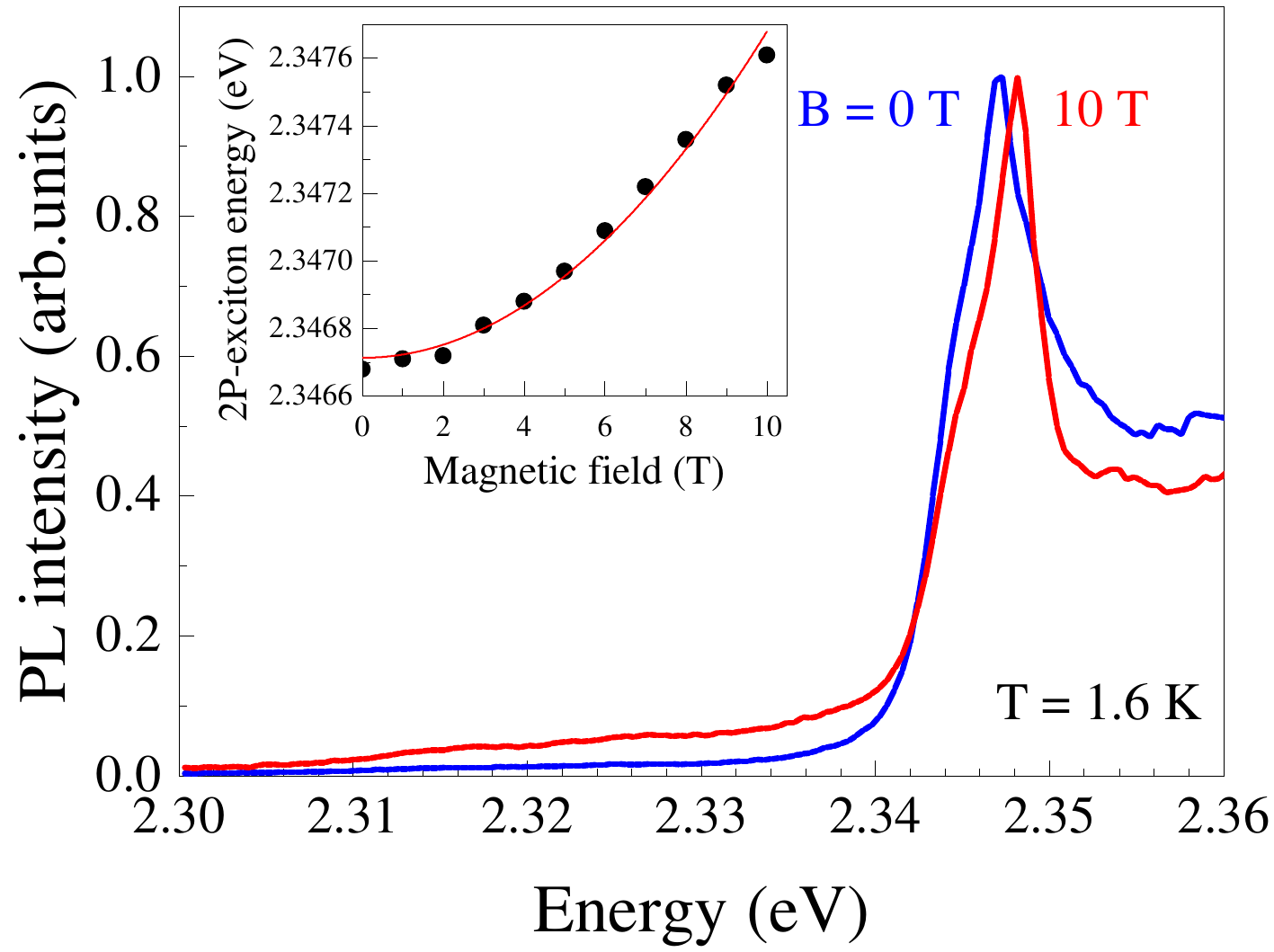}
   \caption{Two-photon PLE spectra of CsPbBr$_3$ in zero magnetic field and in 10~T. The PL detection energy is 2.30~eV. Inset: the magnetic field dependence of the 2P-exciton energy.  Symbols are experimental data. $B^2$ fit with diamagnetic coefficient  $c_d^{2P}=10$~$\mu$eV/T$^2$ is shown by the red line.} \label{fig:3}
\end{figure}

Figure~\ref{fig:3} shows the two-photon PLE spectra of CsPbBr$_3$ measured in zero magnetic field and in 10~T. One can see a pronounced resonance of the 2P exciton with a maximum at $E_{2P}=2.3467$~eV at zero magnetic field. At $B=10$~T the maximum is shifted to 2.3476~eV. The diamagnetic shift of the 2P state is shown in the inset of Fig.~\ref{fig:3}. The red line shows a $B^2$ fit using the diamagnetic coefficient $c_d^{2P}=10$~$\mu$eV/T$^2$. Note, that this state corresponds to the 2P$_{10}$ state with $m=1$ (orbital momentum) and $q=0$ (its projection on the light wave vector) following Ref.~\onlinecite{Makado1986}, where magneto-exciton states are calculated within a hydrogen-like model. 
 
\textbf{Evaluation of exciton parameters.}

We now turn to the evaluation of the fundamental exciton parameters which control its ground and excited Rydberg states: binding energy, reduced mass, and effective dielectric constant. There are several challenges involved: 

(i) The most straightforward way to measure the exciton Rydberg energy, i.e., the binding energy of its 1S state, would be to measure it as $R^*=E_g-E_T$. But in the studied CsPbBr$_3$ crystal, similar to other lead halide perovskite crystals and polycrystalline films~\cite{Yang2017, Baranowski2020AEM}, the free-particle gap $E_g$ has no pronounced feature in optical spectra. It can, in principle, be evaluated by different means, such as by extrapolation to zero field of the magneto-exciton fan chart measured in strong magnetic fields, but the accuracy of this approach is limited. In Ref.~\onlinecite{Yang2017} the 1S exciton ground state binding energy was measured by comparing the energy positions of 1S and 2S exciton states. However, the latter was pronounced only in high magnetic fields and its position in zero field was extrapolated.

(ii) As already mentioned, in lead halide perovskites the screening of the Coulomb interaction may be different for different exciton states, and consequently exciton energies can deviate from the standard hydrogenic Rydberg series.

One approach to estimate the deviation of the binding energies of the exciton states from the hydrogen-like model is based on comparison of the diamagnetic coefficients for the 1S and 2P states. The hydrogen-like model predicts a ratio $c_d^{1S}/c_d^{2P}= 1/6$; see Ref.~\onlinecite{Makado1986} and  Eqs.~\eqref{gamma2}, \eqref{gamma3}. The experimental values that we obtained here are $c_d^{1S}=1.6$~$\mu$eV/T$^2$ and $c_d^{2P}=10$~$\mu$eV/T$^2$. Their ratio is $1/6.25$, which evidences the good validity of the hydrogen-like model for describing excitons in CsPbBr$_3$.  

For accurate and self-consistent evaluation of the exciton parameters we treat the effective dielectric constant $\varepsilon_{\rm eff}$ for 1S ($\varepsilon_{\rm 1S}$) and 2P ($\varepsilon_{\rm 2P}$) exciton states as independent fitting parameters and  determine them separately from the analysis of the experimental data, see SI, S2 for details. We have four parameters to be determined: the exciton reduced mass ($\mu$), the band gap energy ($E_g$), and the effective dielectric constants for the 1S and 2P states ($\varepsilon_{\rm 1S}$ and $\varepsilon_{\rm 2P}$). We use, therefore, four experimentally measured values: the diamagnetic coefficients ($c_d^{\rm 1S}$ and $c_d^{\rm 2P}$) and exciton resonance energies ($E^{\rm 1S}=E_T$, and $E^{\rm 2P}$). We use the four following equations, which link the evaluated and measured parameters. 
\begin{eqnarray}\label{1S_energy}
 E^{1S}=E_g-E_b^{1S}=E_g-\frac{e^4}{2\hbar^2}\frac{\mu}{\varepsilon_{1S}^2} \,, \\
\label{2P_energy}
 E^{2P}=E_g-E_b^{2P}=E_g-\frac{e^4}{8\hbar^2}\frac{\mu}{\varepsilon_{2P}^2} \,, \\
\label{gamma2}
 c_d^{1S}= \frac{\hbar^4}{4 e^2 c^2}\frac{\varepsilon_{1S}^2}{\mu^3}  \,, \\
\label{gamma3}
 c_d^{2P}= \frac{3\hbar^4}{2 e^2 c^2}\frac{\varepsilon_{2P}^2}{\mu^3}  \,. 
\end{eqnarray}

An equal number of evaluated parameters and linear equations provides accurate and unequivocal evaluation of all parameters. The following  parameters are determined:  $\mu=0.18 m_0$, $E_g=2.3545$~eV, $\varepsilon_{1S}=8.66$, and $\varepsilon_{2P}=8.84$. They bring us to the exciton binding energies of $E_b^{1S}=32.5$~meV and $E_b^{2P}=7.8$~meV for the 1S and 2P states, respectively.
Note, that the values of $\varepsilon_{1S}$ and $\varepsilon_{2P}$ are quite close, therefore it is safe to assume that  all the exciton states can be treated within the hydrogen-like model with  a single effective dielectric constant $\varepsilon_{{\rm eff}}\approx 8.7$ with reasonable accuracy.  Moreover, the ratio $E_b^{1S}/E_b^{2P}\approx 4.17$ is close to the value of $4$ that is predicted by the hydrogen-like model (Eqs.~\eqref{Rydberg} and \eqref{EBP2}).  From Eq.~\eqref{Bohr_radius} we get the Bohr radius of the 1S state $a_B^{1S}=2.55$~nm.  
%Note, that we obtained the exciton binding energy very close to one reported in Ref. [\onlinecite{Yang2017}] $E_b^{1S}=33$~meV, although the exciton reduced mass and effective dielectric constant $\mu=0.126 m_0$ and $\varepsilon_{{\rm eff}}=7.3$, correspondingly, are somewhat different. 
%\commentMarina{I removed from here a small piece of text about comparison with literature, as it was duplicated  below.}

In Table~\ref{hh_singlet} we summarize the parameters measured and evaluated in this work, and further details are given in Supporting Information Table S2.

\begin{widetext}
\begin{center}
\begin{table}[hbt]
\caption{Exciton parameters in bulk CsPbBr$_3$ single crystal. $T=1.6$~K.}
\label{hh_singlet}
\begin{tabular}{|p{2cm} |p{3cm} |p{12cm} |}
\hline
 Parameter& Value  & Comments \\
\hline 
 $E_T=E^{1S}$& $2.3220$~eV & Energy of transverse exciton.  \\
 $E_L$& $2.3273$~eV & Energy of longitudinal exciton. \\
 $\hbar \omega_{LT}$& $5.3$~meV & Longitudinal-transverse splitting.  \\
 $\hbar \Gamma$& $6.7$~meV & Exciton damping.  \\
 $4 \pi \alpha_0$& $0.0195$ & Exciton polarizability. \\
 $\varepsilon_b$ & 4.3  & Background dielectric constant. Taken as $\varepsilon_b=\varepsilon_{\infty}=4.3$ from Ref.~\onlinecite{Miyata2017}.\\
\hline 
 $E^{2P}$ & $2.3467$~eV & Energy of the 2P exciton.  \\
\hline 
 $g_X$ & $+2.35$ & Exciton $g$-factor.  \\
 $c_d^{1S}$ & $1.6$~$\mu$eV/T$^2$ & Diamagnetic coefficient of 1S exciton.  \\
 $c_d^{2P}$ & $10$~$\mu$eV/T$^2$  & Diamagnetic coefficient of 2P exciton. \\
\hline
$E_b^{1S}$ & $32.5$~meV & Binding energy of the 1S exciton state, i.e. the exciton Rydberg $R^*$ in hydrogen-like model. \\
$E_b^{2P}$ & $7.8$~meV & Binding energy of the 2P exciton state. \\
$E_g$ & $2.3545$~eV & Band gap energy. \\
$\varepsilon_{\rm eff}$ & $8.7$ & Effective dielectric constant. \\
$\mu$ & $0.18m_0$ & Reduced exciton mass. \\
$M$ & $0.72m_0$ & Translation exciton mass. \\
$a_B^{1S}$ & $2.55$~nm & Bohr radius of the exciton ground state. \\
\hline
\end{tabular}
\end{table}
\end{center}
\end{widetext}

\section*{Discussion}
\label{sec:discussion}

\begin{figure}[hbt]
\includegraphics[width=\columnwidth]{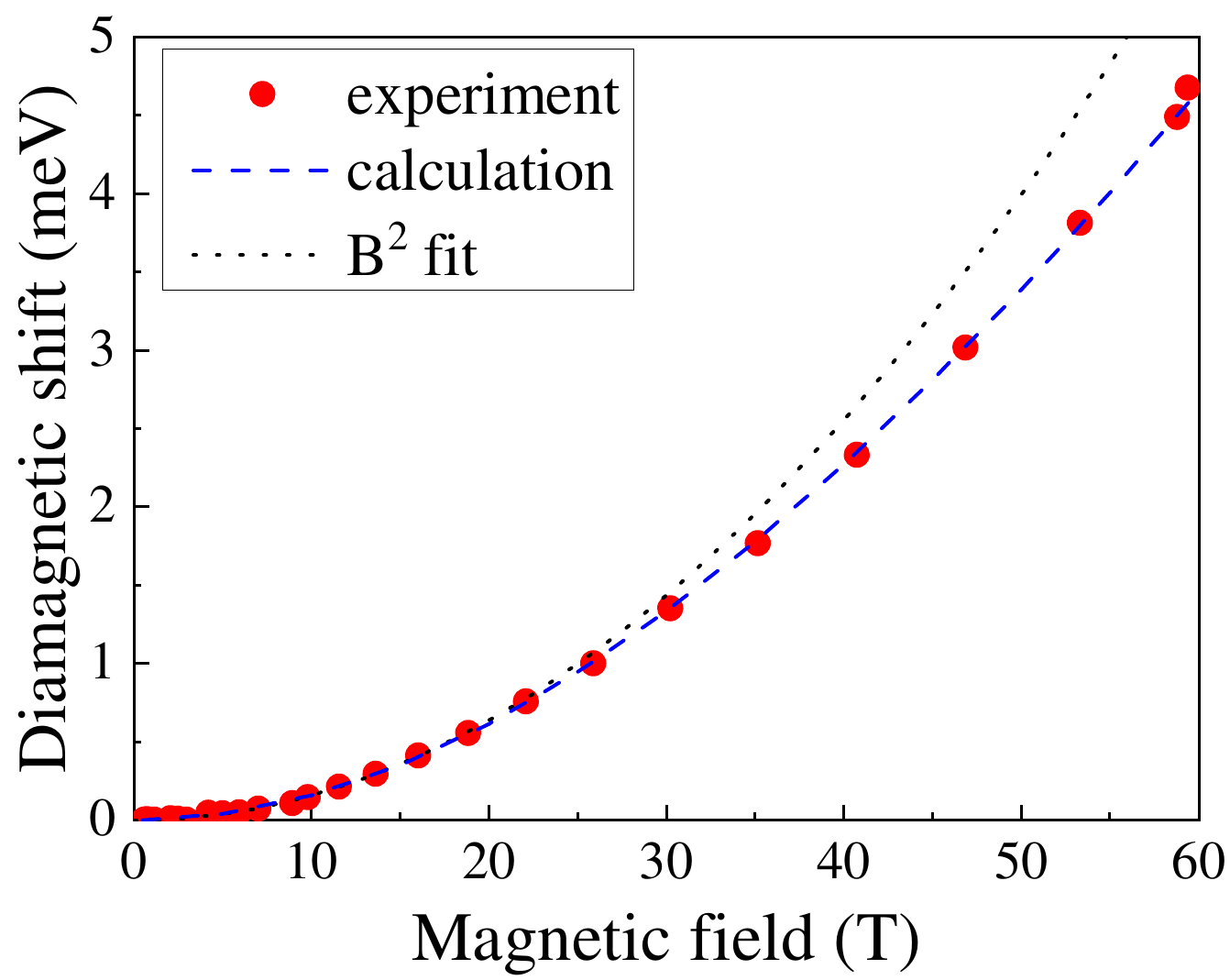}
\caption{The 1S-exciton shift in magnetic field measured (circles), calculated with Eq.~\eqref{exHam} (dashed line) and $B^2$ fit (dotted line). $\mu=0.18 m_0$ and $\varepsilon_{{\rm eff}}= 8.7$.  }
   \label{fig:4}
\end{figure}

Let us have a close look at the field-dependent energy shift of the 1S exciton state, plotted in Fig.~\ref{fig:2}(c) as a function of $B^2$ and also in Figure~\ref{fig:4} as a function of $B$. In both cases one can see that the pure $B^2$ shift is valid primarily in magnetic fields weaker than 25~T. Here, the condition for the diamagnetic exciton is fulfilled with high accuracy. Namely, the cyclotron radius of the charge carriers $\sqrt{c\hbar/eB}$ should be larger than the exciton Bohr radius $a_B$. The critical magnetic field for this criterion is $B_c=c\hbar/(e a_B^2)$. In CsPbBr$_3$ it gives $B_c\approx 100$~T, which is only a factor of $\sim$2 larger than the 60~T fields we use.  Therefore we can expect a small deviation from purely quadratic diamagnetic shifts when approaching  the high fields used in this work.

The dashed line in Fig.~\ref{fig:4} shows the results of a full calculation that accounts for the deviation from the weak magnetic field regime. This simple model Hamiltonian neglects the motion of the exciton center of mass.
\begin{equation}\label{exHam}
\widehat{H}=\frac{\hbar^2}{2\mu}\Delta-\frac{e^2}{\varepsilon_{\text{eff}}r}+\frac{1}{8}\frac{e^2}{c^2}\frac{B^2}{\mu}(x^2+y^2) \, .
\end{equation}
Here $\Delta$ is the Laplace operator, acting on the electron and hole relative coordinate $\bm{r}$, and $x$ and $y$ are components of $\bm{r}$ in the plane perpendicular to the magnetic field. Equation \eqref{exHam} was solved numerically. We used values of $\mu=0.18 m_0$ and $\varepsilon_{{\rm eff}}\approx 8.7$ that were determined above. One can see an excellent agreement with experiment in the whole range of  measured magnetic fields, which justifies our assumptions made for the calculation of exciton parameters and shows the consistency of the obtained set of exciton parameters.

The exciton parameters measured in this study allow us to estimate the Kane energy $E_p$, which is connected with the interband momentum matrix element $P_{cv}$ as $E_p=2P_{cv}^2/m_0$. The Kane energy characterizes the efficiency of the light interaction with the crystal, for details see Supporting Information S3~\cite{Kane,Yu,kirstein2021nc}. We have two ways to do that. In the first approach, we estimate $E_p$ from the electron and hole effective masses \cite{Yu,kirstein2021nc}. In lead halide perovskites the electron and hole effective masses are similar~\cite{Efros}, therefore we estimate $m_h=2\mu$, which for $\mu=0.18  m_0$ gives $m_h=0.36  m_0$. Then taking the band gap energy of CsPbBr$_3$ of  $E_g=2.35$~eV and the conduction band spin orbit splitting $\Delta_{SO}= 1.5$~eV~\cite{kirstein2021nc} we get from Eq.~(S13) $E_p = 11.8-12.5$~eV. The two values results from evaluations via the electron and hole effective masses.

We made a second evaluation of $E_p$ from the link between the longitudinal-transverse splitting and $E_p$ given by \cite{Ivchenko_book_2007}:   
\begin{equation}\label{ELT2}
\hbar\omega_{LT}=\frac{4e^2\hbar^2}{3m_0 E_T^2}\frac{E_p}{\varepsilon_b a_B^3}.
\end{equation}
Using parameters from Table~I, we calculate $E_p=15$~eV, which is close to the value from the first evaluation. This agreement justifies use of the exciton-polariton approach and obtained exciton parameters. Note that due to the difference of the band structure of the lead halide perovskites and conventional zinc-blende semiconductors, for the latter in Eq.~\eqref{ELT2} the coefficient 2 should be used instead of 4/3. 

The following evaluations of $E_p$ can be made on the basis of literature data. In Ref.~[\onlinecite{kirstein2021nc}] the effective masses of electrons and holes were calculated by DFT and ETB methods, the reported values are $m_e=0.3 (0.291) m_0$ and $m_h=0.26(0.298)$ for DFT (ETB), respectively. Also the  experimental value for hole effective mass $m_h=0.26 m_0$ was measured by angle-resolved photoelectron spectroscopy in Ref.~[\onlinecite{Puppin2020}]. They correspond to a reduced exciton mass $\mu=0.14-0.15$ and a Kane energy $E_p=13.5-17.5$~eV. In Ref.~[\onlinecite{kirstein2021nc}] from the modeling of the universal dependence of the charge carrier $g$-factors on the band gap energy, the parameter  $P_{cv}=\hbar p/m_0=6.8$~eV\AA ~was evaluated. It corresponds to a Kane energy $E_p=12.1$~eV.  

Magneto-optical experiments in high fields up to 70~T were reported for CsPbBr$_3$ polycrystalline films in Ref.~\onlinecite{Yang2017}. The following exciton parameters were evaluated: $R^*=E_b^{1S}=33 \pm 1$~meV, $\mu = (0.126 \pm 0.01)m_0$, and $\varepsilon_{eff}=7.8$.  The exciton binding energy of the 1S state, estimated as the difference between the energy of the 1S exciton and extrapolation of the magneto-exciton fan chart to get $E_g$, is very close to our value of 32.5~meV. While the reduced exciton mass, evaluated from the fit of the fan chart in strong magnetic fields, is smaller than our determination of $\mu = 0.18m_0$, which is based on diamagnetic shifts in fields below 25~T, it can be measured with high accuracy due to the narrow exciton lines in bulk crystals. The $\varepsilon_{eff}$ evaluated in Ref.~\onlinecite{Yang2017} from $E_b^{1S}$ and $\mu$  is smaller than our value of 8.7, which is related to the smaller $\mu$ value. 
% It is worth to note, that different calculated set of effective masses are reported to be $m_e=0.15 (0.134) m_0$ and $m_h=0.14 (0.128) m_0$ in Refs.~[\onlinecite{Protesescu2015}] and [\onlinecite{Efros}]. In Ref. [\onlinecite{kirstein2021nc}]  electron and hole effective masses were calculated by DFT (ETB), reported values are $m_e=0.3 (0.291)m_0$  and $m_h=0.26 (0.298)m_0$, which gives reduced exciton  mass about $\mu=0.14\div0.15 m_0$.  . 
The band gap  energy of $E_g=2.3545$~eV that we determine in our study is in line with  the literature data~\cite{Kulbak2015,Hoffman,Yang2017}.

%%%%%%%%%%%%%%%%%%%%%%%%%%%%%%%%%%%%%%%%%%%%%%%%%%%%%%%%%%%%%%%%%%%%%%%%%%%%%%%%%%

\section*{Conclusions}
\label{sec:conclusions}
Magneto-optics and two-photon spectroscopy of CsPbBr$_3$ crystals allow us to determine a detailed set of exciton and exciton-polariton parameters  summarized in Table~\ref{hh_singlet}. The combination of these approaches can be readily extended to other lead halide perovskites and their nanostructures.

\section*{Methods}

\textbf{Samples.} 
The CsPbBr$_3$ crystals were grown with a slight modification of the inverse temperature crystallization technique, see Ref.~\cite{Dirin2016}. First, CsBr and PbBr$_2$ were dissolved in dimethyl sulfoxide. Afterwards a cyclohexanol in N,N-dimethylformamide solution was added. The resulting mixture was heated in an oil bath to $105^\circ$C whereby slow crystal growth occurs. The obtained crystals were taken out of the solution and quickly loaded into a vessel with hot ($100^\circ$C) N,N-dimethylformamide. Once loaded, the vessel was slowly cooled down to about $50^\circ$C. After that, the crystals were isolated, wiped with filter paper and dried. The obtained rectangular-shaped CsPbBr$_3$ is crystallized in the orthorhombic modification. The crystals have one selected (long) side along the $c$-axis [001] and two nearly identical sides along the $[\bar{1}10]$ and [110] axes~\cite{feng2020}. 

%The size of the crystal is $\approx3\times2\times7$~mm$^3$. \cD{check whether we need sizes here. I guess we measured another piece in Los Alamos. And in fact we do not controlled its orientation.} \cS{we did use a smaller piece. Maybe we don't need to state sample dimensions}

\textbf{Reflectivity in pulsed magnetic fields.}
The sample was mounted on a custom fiber-coupled probe in a helium bath cryostat with a long tail extending into the bore of a 65~T pulsed magnet. The experiments were performed at a temperature of $T=1.6$~K, with the sample immersed in superfluid helium. Broadband white light from a halogen lamp was coupled down a 100 $\mu$m diameter multimode optical fiber, and light reflected from the sample was collected by a 600-$\mu$m diameter fiber. The light wavevector $\mathbf{k}$ was perpendicular to the sample surface and parallel to $B$ (Faraday geometry).  Thin film circular polarizers were used to select $\sigma^-$ and $\sigma^+$ polarized light. Full optical spectra were acquired every 1~ms continuously throughout the magnet pulse using a fast charge-coupled-device (CCD) camera combined with a 0.3-meter spectrometer. To switch between $\sigma^-$ and $\sigma^+$ polarization, the direction of the magnetic field was switched. Further details of these methods can be found in Ref.~\onlinecite{Stier2016}.

\textbf{Two-photon excitation of photoluminescence.}  In order to obtain information on the energy of the 2P exciton state in absorption, we measured  photoluminescence excitation (PLE) spectra. In this technique the emission was detected at the low energy tail of the PL spectrum and the laser photon energy was tuned across the exciton absorption spectral range. For excitation a pulsed laser system with an optical parametric amplifier (OPA) was used. Its photon energy was tunable in the spectral range of $0.5-4.0$~eV.  The laser pulses had a duration of $2.5$~ps, a linewidth of $0.23$~nm, and a repetition frequency of $30$~kHz. For the two-photon PLE the laser was tuned in the range of $1.15-1.20$~eV. The PL was measured at 2.30~eV, about 50~meV below the exciton energy. For this, it was spectrally selected  by a 10~nm wide bandpass filter and  a spectrometer. For these measurements the sample was placed in the magneto-optical cryostat with direct optical access. The sample was in direct contact with pumped liquid helium at $T=1.6$~K. Static magnetic fields up to 10~T were generated by a split-coil superconducting magnet and applied in the Faraday geometry.

\textbf{ASSOCIATED CONTENT}

\textbf{Supporting Information.}
Additional information on the band structure and exciton fine structure in CsPbBr$_3$. Theoretical consideration of the exciton-polaritons in reflectivity spectra. Table with detailed comments on evaluation of exciton parameters.

\textbf{AUTHOR INFORMATION}

{\bf Corresponding Author} \\
Dmitri R. Yakovlev,  Email: dmitri.yakovlev@tu-dortmund.de\\

\textbf{ORCID}\\
Dmitri R. Yakovlev:   0000-0001-7349-2745\\
Scott A. Crooker:     0000-0001-7553-4718\\
Marina A. Semina:     0000-0003-3796-2329\\
Janina Rautert:       0000-0002-9908-4851 \\
Johannes Mund:        0000-0002-8022-7584\\
Dmitry N. Dirin:      0000-0002-5187-4555\\
Maksym V. Kovalenko:  0000-0002-6396-8938\\
Manfred Bayer:        0000-0002-0893-5949 \\

%\section{Data availability}

%\section{Author contributions}
%D.R.Y., S.A.C., J.R., and J.M. performed the measurements under the guidance of M.B., M.A.S., and D.R.Y. analyzed and  interpreted  the  data. D.N.D. and M.V.K. synthesized the perovskite crystals.  D.R.Y., M.A.S., and S.A.C. wrote the manuscript in close consultation with M.B.

%\section{Additional information}

{\bf Notes}\\
The authors declare no competing financial interests.

{\bf Acknowledgments}\\
%Authors are thankful to V. F. Sapega and I. V. Kalitukha for their contribution to an initial stage of this study. 
The authors are thankful to M. M. Glazov and E. L. Ivchenko for fruitful discussions and to  P. Sercel for valuable advice.
The authors acknowledge financial support of the Deutsche Forschungsgemeinschaft through the Collaborative Research Center TRR142 (Project A11) and the Priority Programme SPP2196 (Project YA 65/26-1).  M.A.S. acknowledges support of the Russian Science Foundation (project 23-12-00300). The National High Magnetic Field Laboratory is supported by the National Science Foundation DMR-1644779, the State of Florida, and the US Department of Energy. S.A.C. acknowledges support from the US Department of Energy ``Science of 100~T'' program.  Work at ETH Z\"urich (D.N.D. and M.V.K.) was financially supported by the Swiss National Science Foundation (grant agreement 186406, funded in conjunction with SPP219 through DFG-SNSF bilateral program) and by ETH Z\"urich through ETH+ Project SynMatLab.

\begin{figure}[hbt]
\includegraphics[width=8 cm] {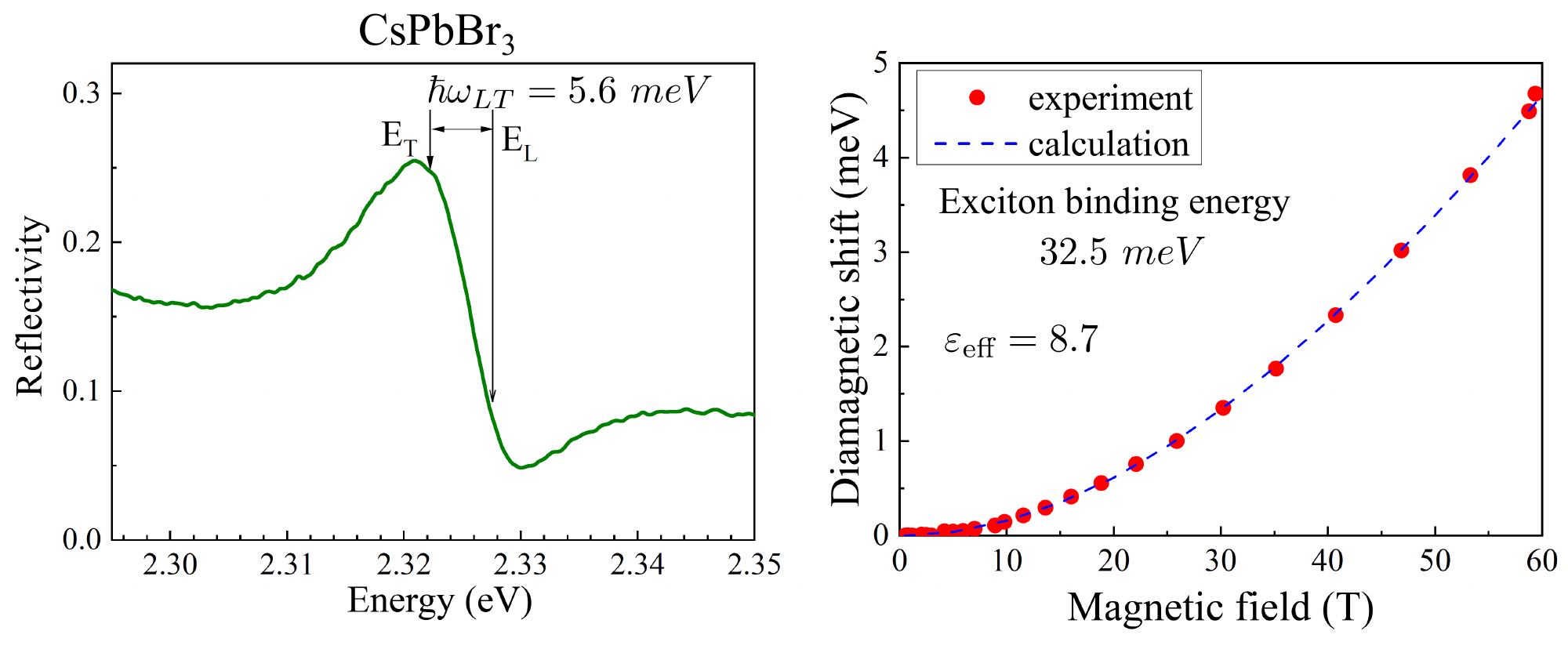}
\caption{TOC figure. } 
%\label{fig:TOC}
\end{figure}

%%%%%%%%%%%%%%%%%%%%%%%%%%%%%%%%%%%%%%
%%%%%%%%%%%%%%%%%%%%%%%%%%%%%%%%%%%%%%

%\section{OPEN QUESTIONS}

%\cD{Below are open questions to discuss and check out before submission.}

% (i) Marina will check whether we should care about spectral shift between $E_T$ that we evaluate from the fitting exciton-polariton reflectivity and maximum in absorption spectrum.  One should check here Kramers-Kronig transformations from Reflectivity to absorption. It seems that the absorption maximum should be close to $E_T$, but would be good to check for possible deviations. Especially if later we will deal with materials with even stronger exciton-polariton. \commentMarina{did not do it yet}

%(ii) Check what Toulouse colleagues means under 2P(1,0) state. They start to observed it at finite magnetic fields and refer to Ref.~\cite{Makado1986} for calculations. This state is shifted considerably faster than the 2S state. It seems to me that it is not what we identify as 2P peak in TPA. We need to be sure that diamagnetic shift of our 2P state in TPA is same as for the 2S state. (Marina will do that). \cD{seems that this is not correct!}\commentMarina{In previous versions there were a mistake with expressions for diamagnetic shits. Now it is corrected.}

%%%%%%%%%%%%%%%%%%%%%%%%%%%%%%%%%%%%%%%%%%%%%%%%%%%%
\clearpage

%%%%%%%%%% Merge with supplemental materials %%%%%%%%%%
%%%%%%%%%% Prefix a "S" to all equations, figures, tables and reset the counter %%%%%%%%%%
\setcounter{equation}{0}
\setcounter{figure}{0}
\setcounter{table}{0}
\setcounter{page}{1}
%\makeatletter
\renewcommand{\theequation}{S\arabic{equation}}
\renewcommand{\thefigure}{S\arabic{figure}}
\renewcommand{\thepage}{S\arabic{page}}
\renewcommand{\thetable}{S\arabic{table}}
%\renewcommand{\thesection}{S\arabic{section}}
%%%%%%%%%% Prefix a "S" to all equations, figures, tables and reset the counter %%%%%%%%%%

\begin{widetext}
\begin{center}

\section*{Supporting Information}

\textbf{\large Exciton-polaritons in CsPbBr$_3$ crystals revealed by optical reflectivity in high magnetic fields and two-photon spectroscopy}\\

\vspace{\baselineskip}

 \textit{Dmitri R. Yakovlev, Scott A. Crooker, Marina A. Semina, Janina Rautert, Johannes Mund, Dmitry N. Dirin, Maksym V. Kovalenko, Manfred Bayer
}
\end{center}

\subsection*{S1. Band structure and excitons in orthorhombic CsPbBr$_3$}

Bulk CsPbBr$_3$ can exist in three structural crystalline phases: the high symmetry cubic ($O_h$ point symmetry) phase, which is stable at temperatures higher than $133^{\circ} \text C$; the tetragonal ($D_{4h}$)  phase is stable between $133^{\circ} \text C$ and $88^{\circ} \text C$; and the orthorhombic ($D_{2h}$) phase stable below  $88^{\circ} \text C$~\cite{Stoumpos2013_SI}. In the tetragonal and orthorhombic phases the direct band transitions are realized at the $\Gamma$ point of the Brillouin zone, while in the cubic phase the direct band transition is at the $R$ point of the Brillouin zone (corner of the cube). 

The band structure of CsPbBr$_3$ is ``inverted'' in comparison to conventional III-V and II-VI semiconductors such as GaAs or ZnSe~\cite{Even2012_SI, kirstein2021nc_SI}.  The conduction band in CsPbBr$_3$ consists of two subbands having net angular momentum $J_e=3/2$ and $J_e=1/2$, while the valence band is characterized solely by spin $J_h=1/2$. The valence band is two-fold degenerate by spin and originates mostly from the overlap of the Pb metal $6s$-orbitals and the Br $4p$-orbitals. Spin-orbit coupling does not affect the valence band, although for the lower symmetry in the tetragonal and orthorhombic phases the reduction of symmetry results in effective crystal fields contributing to the valence band dispersion.  The conduction band is formed  mostly by the Pb  $6p$-orbitals and it is dominated by strong spin-orbit coupling due to the rather large atomic weight of Pb. This leads to a splitting of the conduction band into $J_e=3/2$, which is split by crystal field in tetragonal and orthorhombic phases, and $J_e=1/2$ subbands. The lowest conduction subband is the split-off $J_e=1/2$ subband. A sketch of the band structure of CsPbBr$_3$ is shown in Fig.~\ref{fig:band_SI}a. In this figure $E_g$ is the band gap energy, $\Delta_{le}$ and $\Delta_{he}$ are the splittings between bottom conduction band and the bands of light and heavy electrons (\textit{le} and \textit{he}), respectively.  

\begin{figure}[hbt]
\includegraphics[width=0.75\columnwidth]{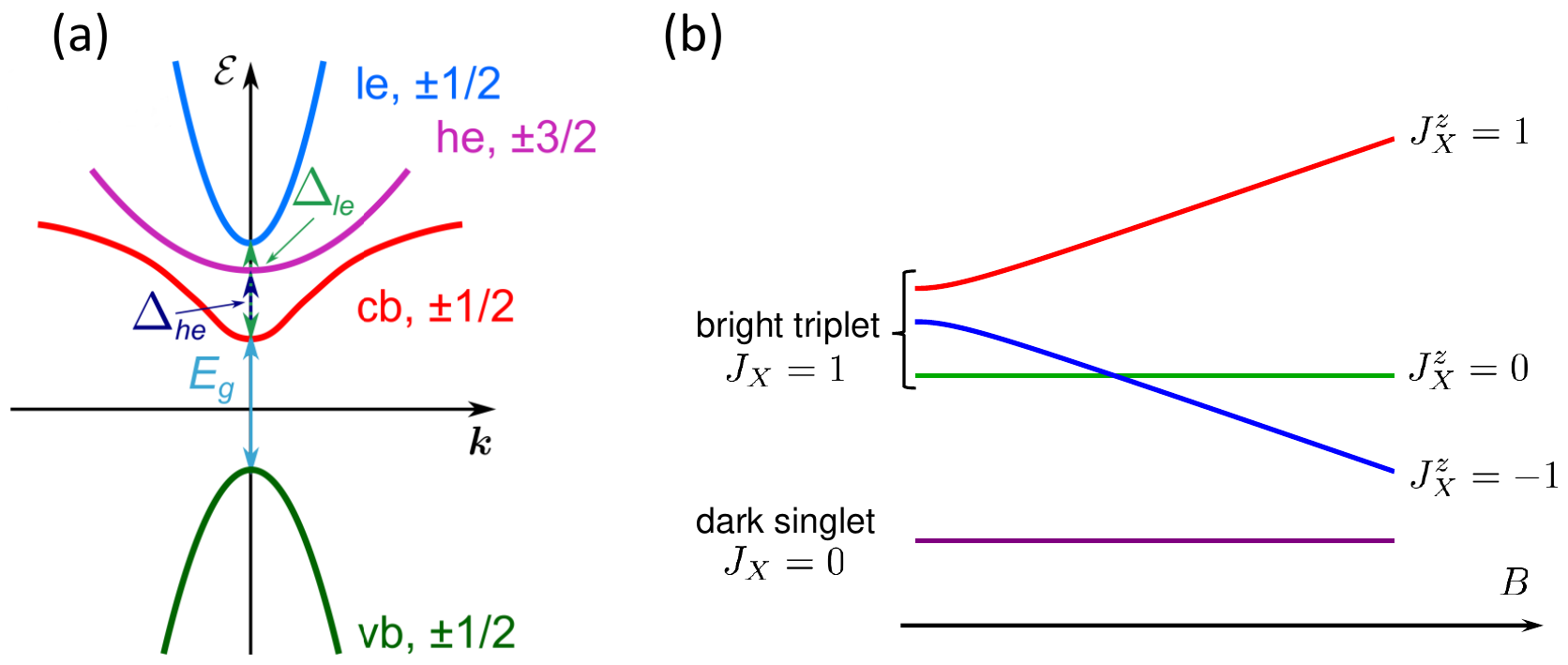}
\caption{(a) Schematics of the band structure of CsPbBr$_3$ in the orthorhombic phase. (b) The scheme of exciton levels in magnetic field. }
   \label{fig:band_SI}
\end{figure}

The band edge excitons in CsPbBr$_3$ are  formed by the Coulomb interaction between a hole with $J_h=1/2$ at the top of the valence band and an electron with $J_e=1/2$ at the bottom of the lowest conduction subband. The exciton ground  state (which in a hydrogen-like model would have S-like envelope function of electron-hole relative motion), is split by the electron-hole exchange interaction and the Rashba effect into a singlet dark state with net angular momentum $J_X=0$, and an  optically active triplet state with $J_X=1$. The lowest exciton state is the dark singlet state.  The triplet state is fully split in the lower symmetry orthorhombic phase by the spin projection $J_X^z$ on the $c$-axis direction ($[001]$). Two of them with $J_X^z=\pm 1$ are active in the Faraday geometry with magnetic field being directed along the $[001]$ direction~\cite{SBaranowski2019_SI}. The  exciton fine structure in bulk CsPbBr$_3$ has not been measured yet. The data for bulk MAPbBr$_3$ shows values $\sim 200$ $\mu$eV for the bright exciton fine structure splitting~\cite{SBaranowski2019_SI}  and we assume similar values for CsPbBr$_3$.  The scheme of exciton levels in applied magnetic field is shown in Fig.~\ref{fig:band_SI}b. The order of exciton states within the bright triplet and the relative energy splittings are arbitrary.  The magnitudes of Zeeman splitting in the magnetic fields used in this work are larger than the assumed exciton fine structure splittings, so that states forming bright triplets would be effectively mixed, and the observed linewidths of reflectivity spectra are quite large also. Thus, we could not resolve the exciton fine structure in our experiments and we do not expect to have different reflectivity spectra for various crystal orientations. As it was shown in Ref.~[\onlinecite{kopteva2023_gX_SI}], the exciton $g$-factor in CsPbBr$_3$ is almost isotropic and our neglecting of exciton fine structure while fitting experimental data is justified.

\subsection*{S2. Modeling of exciton-polaritons in reflectivity}

Exciton-polaritons are quasiparticles created by  light-matter interaction that couples excitons with photons. In semiconductors with sufficiently strong light-matter interaction such as CsPbBr$_3$, exciton-polaritons determine the optical properties in the vicinity of the exciton resonance. Their properties can be studied by measuring reflectivity spectra in this spectral range.  The dispersion relation is:
\begin{equation}\label{polariton}
\frac{c^2k^2}{\omega^2}=\varepsilon(\omega, k),
\end{equation}
where $\omega$ is the incident light frequency and ${\bm k}$ is the light wave vector. The exciton-polaritons are characterized by the dependence of the dielectric function $\varepsilon(\omega, k)$ not only on the light frequency, but also on the wave vector, i.e. they exhibit spatial dispersion. Schematically the exciton-polariton dispersion  is shown in Fig.~\ref{fig:1}(c) in the main text. Incident light with frequency $\omega$ gives rise to one or two  polaritons (depending on $\omega$) in the semiconductor crystal. Consequently, the boundary conditions for electromagnetic waves are insufficient to solve  Maxwell's equations. One can, therefore, either solve the non-local Maxwell equations~\cite{Maradudin_SI,Agarwal_SI} or introduce additional boundary conditions~\cite{Hopfield_SI,Sell_SI}. For our study it is sufficient to use simple boundary conditions in the form of a ``dead'' layer~\cite{Hopfield_SI}, a layer with thickness $L$ near a semiconductor surface in which excitons can not exist. 

Here we consider a normal incidence of light wave on a semi-infinite dielectric crystal. Near the exciton-polariton resonance if the electric field and exciton polarization are parallel or perpendicular  to the light wave vector (for uniaxial crystals electric field has to be perpendicular or parallel to optical axis), the dielectric function in  Eq.~\eqref{polariton} takes the form~\cite{Hopfield_SI}:
\begin{equation}\label{epsilon_1}
\varepsilon(\omega, k)=\varepsilon_{b}+\frac{4\pi\alpha_0 E_T^2}{E_T^2-\hbar^2 \omega^2+\frac{\hbar^2 k^2}{M}E_T-{\mathrm i} \hbar^2 \omega \Gamma},
\end{equation}
where $E_T$ is the energy of the transverse (``mechanical") exciton state in the absence of the spatial dispersion and with the resting center of masses, and $\bm{k}$ is the exciton center of masses wave vector. Due to momentum conservation, exciton center of masses wave vector $\bm k$ is equal to the wave vector of incident light. $M=m_e+m_h$ is the  translation exciton mass with $m_{e(h)}$ being the electron (hole) effective mass. Note, that the electron and hole relative motion in the exciton is described by the reduced exciton mass $\mu$, defined as $\mu^{-1}=m_e^{-1}+m_h^{-1}$. The parameter $\varepsilon_{b}$ is the background dielectric constant at the energy of the exciton-polariton resonance, here we take it as a high frequency dielectric constant of the crystal. The parameter $4\pi\alpha_0$ is the exciton polarizability, which controls the exciton oscillator strength, and $\Gamma$ is the damping. Parameter $\varepsilon_{b}$ is the background dielectric constant on the frequency of excitonic transition without taking into account the exciton resonance.  Note, that in literature on exciton-polaritons the background dielectric constant $\varepsilon_{b}$ is sometimes labeled as $\varepsilon_{0}$, see e.g. Refs.~[\onlinecite{Hopfield_SI}] and [\onlinecite{Maradudin1973_SI}]. But this $\varepsilon_{0}$ should not be mixed with the stationary dielectric constant at zero frequency, as rather $\varepsilon_{b} \equiv \varepsilon_{0} \approx \varepsilon_{\infty}$.

The exciton polarizability describes the efficiency of light-matter interactions resulting in mixing of the transverse and longitudinal modes of the exciton-polariton. It is connected with the longitudinal-transverse splitting energy $\hbar\omega_{LT}$ as~\cite{Hopfield60_SI,Sell_SI}:   
\begin{equation}\label{ELT1}
\hbar\omega_{LT}=\frac{4\pi \alpha_0 E_T}{2\varepsilon_{b}}.
\end{equation} 
Then,
\begin{equation}\label{epsilon}
\varepsilon(\omega, k)=\varepsilon_{b} \left(1+\frac{2\hbar\omega_{LT} E_T}{E_T^2-\hbar^2 \omega^2+\frac{\hbar^2 k^2}{M}E_T-{\mathrm i} \hbar^2 \omega \Gamma} \right).
\end{equation}
Solving the dispersion relation of exciton-polaritons, Eq.~\eqref{polariton}, where $\varepsilon(\omega, k)$  is defined in Eq.~\eqref{epsilon_1}, we find the refractive index of the media. The resulting reflection coefficient is \cite{Hopfield60_SI}:
\begin{equation}\label{refl}
R=\frac{1-n^*}{1+n^*},
\end{equation}
where the boundary conditions are taken into account in the form of the ``dead'' layer
\begin{equation}\label{nn}
n^*=\sqrt{\varepsilon_b}\left[\frac{(n^{\dagger}+\sqrt{\varepsilon_b})e^{-2\mathrm i k \sqrt{\varepsilon_b} L}-\sqrt{\varepsilon_b}+n^{\dagger}}{(n^{\dagger}+\sqrt{\varepsilon_b})e^{-2\mathrm i k \sqrt{\varepsilon_b} L}+\sqrt{\varepsilon_b}-n^{\dagger}}\right], \quad n^{\dagger}=\frac{n_1n_2+\varepsilon_b}{n_1+n_2},
\end{equation} 
where $L$ is the thickness of the dead layer, and $n_1$ and $n_2$ are refractive indices found as roots of the dispersion relation \eqref{polariton} and corresponding to the two lower and upper exciton-polariton branches. Substituting Eq.~\eqref{nn} into Eq.~\eqref{refl} we model the reflectivity spectra in the vicinity of the exciton-polariton resonance, which we treat as isolated and do not take into account possible overlap between different exciton states in the spectrum.   

In our simplified model, the damping $\Gamma$ describes the combined contribution of both homogeneous and inhomogeneous broadening. We assume that homogeneous broadening dominates in our case due to the high quality of the CsPbBr$_3$ crystal under study.   We also neglect the complex conduction band structure, which is justified by the very large value of spin-orbit splitting $\Delta_{SO}=1.5$~eV, and treat both the topmost valence band and the bottom conduction as isotropic and parabolic.  

To fit the reflectivity spectrum in the vicinity of the exciton-polariton resonance we take $M$, $E_T$, $\hbar\omega_{LT}$, $\Gamma$, and $L$ as the fitting parameters. The longitudinal exciton energy is calculated as $E_L=E_T+\hbar\omega_{LT}$. In calculations we take the background dielectric constant $\varepsilon_{b}=\varepsilon_{\infty}=4.3$ from Ref.~[\onlinecite{Miyata2017_SI}].

\begin{figure}[hbt]
\includegraphics[width=0.55\columnwidth]{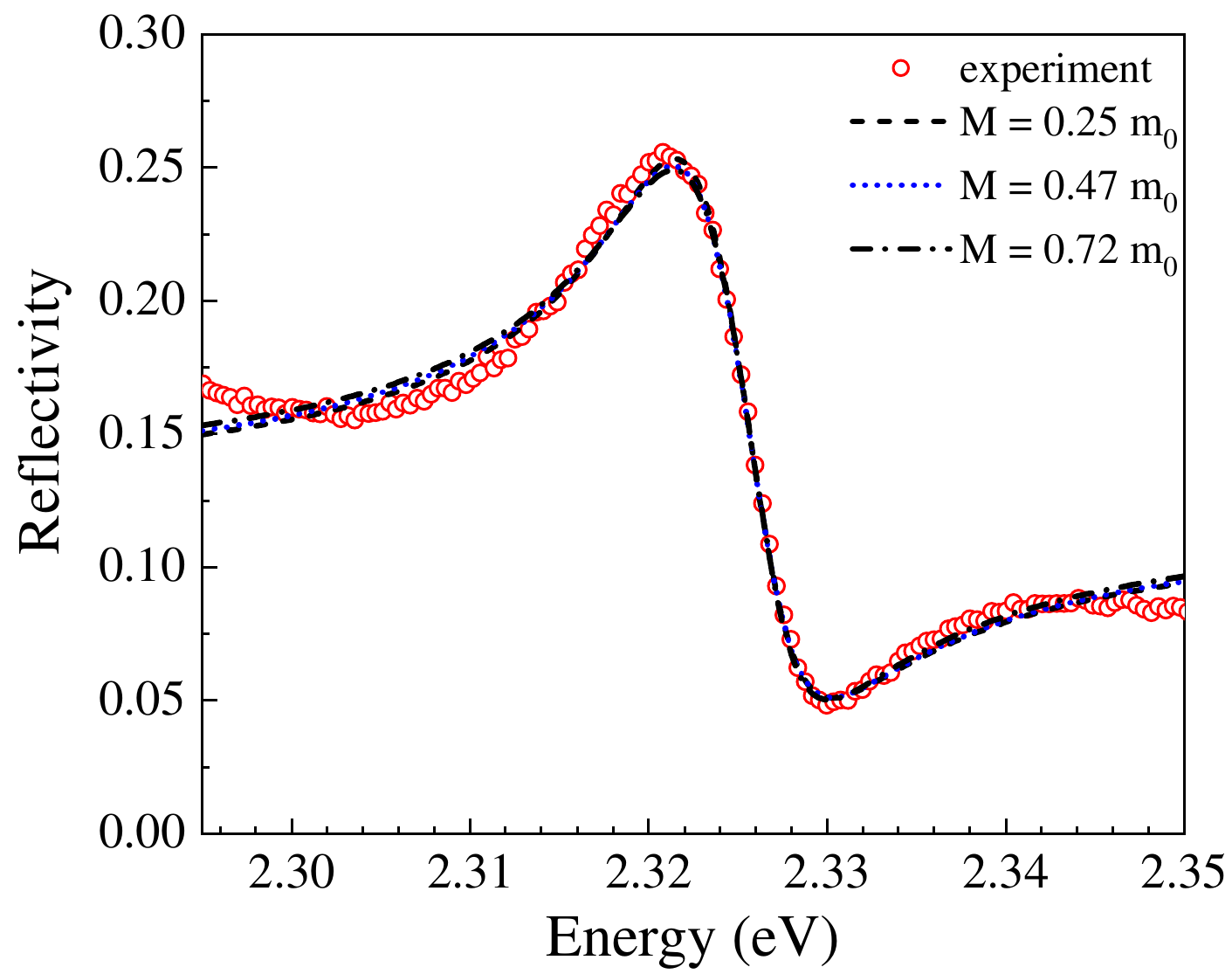}
\caption{Modeling of the exciton-polariton resonance measured in reflectivity (open circles)  for a CsPbBr$_3$ crystal. Lines show calculation results using exciton translation masses of $M=0.25 m_0$, $0.47 m_0$, and $0.72 m_0$.   }
   \label{fig:5}
\end{figure}

\begin{center}
\begin{table}[hbt]
\caption{Exciton-polariton parameters for CsPbBr$_3$ at $T=1.6$~K as determined by modeling of the reflectivity spectrum in Figure~\ref{fig:5}.  }
\label{fit_parameters_SI}
\begin{tabular}{|p{1.5cm} |p{1.5cm} |p{2cm} |p{1.5cm}  |p{1.5cm}  | }
\hline
$M$ & $E_T$ (eV)  & $\hbar\omega_{LT}$ (meV) & $\Gamma$ (meV) & $L$ (nm) \\
\hline 
$ 0.25 m_0$ & 2.3219  & 5.4& 6.2 & 3.1 \\

\hline 
$ 0.47 m_0$ & 2.322  & 5.4 & 6.8 & 3.5 \\

\hline 
$ 0.72 m_0$ & 2.322  & 5.3 & 6.7 & 2.3 \\

\hline
\end{tabular}
\end{table}
\end{center}

The results  of the fitting of reflectivity spectra in zero magnetic field with different translation exciton masses, $M$, are shown by lines in Fig.~\ref{fig:5}. For each fit we fixed $M$ and varied other parameters. The best fit parameters are given in Table~\ref{fit_parameters_SI}. One can see that the exciton parameters are only weakly sensitive to the choice of $M$, which was varied from  $0.25 m_0$ up to $0.72 m_0$. The value of  $M=0.47m_0$ is in line with Refs.~\onlinecite{Stoumpos2013_SI,Yang2017_SI, Yettapu_SI}.  Solving equations \eqref{1S_energy}, \eqref{2P_energy}, \eqref{gamma2}, and \eqref{gamma3}  from the main text we  unambiguously found the exciton  reduced  mass  $\mu=0.18 m_0$. Then, considering that in the lead halide perovskites the electron and hole effective masses are close to each other, we get  $M \approx 4\mu = 0.72 m_0$.  Therefore, for all fits  in various magnetic fields we use $M=0.72 m_0$. At zero magnetic field it gives us the following exciton-polariton parameters: $E_T=2.322$~eV, $\hbar\omega_{LT}=5.3$~meV, $\hbar\Gamma =6.7$~meV, and $L=2.3$~nm. The obtained dead layer thickness, although it is only a model parameter, is of similar magnitude as the exciton Bohr radius, which seems reasonable.

Note that the energy of transverse exciton, $E_T$, which is crucial for determination of the exciton parameters,  shows almost no dependence on the exciton translation mass, and on other parameters showed only very slight dependence. Also, although we have several fitting parameters, most of them can be determined with a high accuracy, as they are individually responsible for specific aspects of the reflectivity spectrum: its spectral energy position, amplitude, shape,  and line broadening (damping). Therefore, the obtained values of exciton parameters are reliable and are evaluated with high accuracy. 

In Table~\ref{hh_singlet_SI} we summarize parameters measured and evaluated for  CsPbBr$_3$ in this paper and give comments on how they were obtained.

%\begin{widetext}
\begin{center}
\begin{table}[h]
\caption{Exciton parameters in bulk CsPbBr$_3$ crystal. $T=1.6$~K.}
\label{hh_singlet_SI}
\begin{tabular}{|p{2cm} |p{3cm} |p{12cm} |}
\hline
 Parameter& Value  & Comments \\
\hline 
 $E_T=E^{1S}$& $2.3220$~eV & Energy of transverse exciton. Obtained from the fit of reflectivity spectra. \\
 $E_L$& $2.3273$~eV & Energy of longitudinal exciton. Calculated by expression $E_L=E_T+\hbar\omega_{LT}$. \\
 $\hbar \omega_{LT}$& $5.3$~meV & Longitudinal-transverse splitting. Obtained from the fit of reflectivity spectra at zero magnetic field. \\
 $\hbar \Gamma$& $6.7$~meV & Exciton damping. Obtained from the fit of reflectivity spectra at zero magnetic field. \\
 $4 \pi \alpha_0$& $0.0195$ & Exciton polarizability. Calculated from Eq.~\eqref{ELT1}.\\
 $\varepsilon_b$ & 4.3  & Background dielectric constant. Taken as $\varepsilon_b=\varepsilon_{\infty}=4.3$ from Ref.~\onlinecite{Miyata2017_SI}.\\
\hline 
 $E^{2P}$ &  2.3467 & Energy of 2P exciton. Measured by two-photon absorption. \\
\hline 
 $g_X$ & $+2.35$ & Exciton $g$-factor. Obtained from experimentally measured Zeeman splitting of exciton, namely of  $E_T$ splitting in $\sigma^+$ and $\sigma^-$ polarizations.  \\
 $c_d^{1S}$ & $1.6$~$\mu$eV/T$^2$ & Diamagnetic coefficient of 1S exciton. Obtained from fitting the center-of-gravity of dependence $E_T(B)$ in $\sigma^+$ and $\sigma^-$ polarizations. \\
 $c_d^{2P}$ & $10$~$\mu$eV/T$^2$  & Diamagnetic coefficient of 2P exciton. Obtained by fitting the energy shift of 2P exciton state in magnetic field measured by two-photon absorption. \\
\hline
$E_b^{1S}$ & $32.5$~meV & Binding energy of 1S exciton state, i.e. the exciton Rydberg $R^*$ in the hydrogen-like model. Calculated as $E_b^{1S}=E_g-E^{1S}$.  \\
$E_b^{2P}$ & $7.8$~meV & Binding energy of 2P exciton state. Calculated as $E_b^{2P}=E_g-E^{2P}$.\\
$E_g$ & $2.3545$~eV & Band gap energy. Evaluated from experimental data. \\
$\varepsilon_{\rm eff}$ & $8.7$ & Effective dielectric constant. Evaluated from experimental data.\\
$\mu$ & $0.18m_0$ & Reduced exciton mass. Evaluated from experimental data. \\
$M$ & $0.72m_0$ & Translation exciton mass. Evaluated as $M=4\mu$ suggesting that $m_e=m_h$. \\
$a_B^{1S}$ & $2.55$~nm & Exciton Bohr radius. Evaluated from Eq.~\eqref{Bohr_radius} using experimental data. \\
\hline
\end{tabular}
\end{table}
\end{center}
%\end{widetext}

\subsection*{S3. Longitudinal-transverse splitting of excitons in lead halide perovskites }

The valence band of lead halide perovskites is S-like (as are the conduction bands of III-V and II-IV semiconductors). The bottom conduction subband is an analog of spin-split valence subband of conventional semiconductors.
In the quasicubical approximation the basis functions of the valence and conduction bands in perovskites are \cite{kirstein2021nc_SI, Yu_SI}: 
\begin{equation}
u_{v,+\frac{1}{2}}=\mathrm i S\uparrow,\quad u_{v,-\frac{1}{2}}=\mathrm i S\downarrow,
\end{equation}
\begin{equation} 
u_{c, +\frac{1}{2}}=-\frac{1}{\sqrt{3}}Z\uparrow-\frac{1}{\sqrt{3}}(X-\mathrm i Y)\downarrow,\quad u_{c, -\frac{1}{2}}=\frac{1}{\sqrt{3}}Z\downarrow-\frac{1}{\sqrt{3}}(X+\mathrm i Y)\uparrow
\end{equation}
while basis functions of heavy electrons (analog of heavy holes) are \cite{kirstein2021nc_SI, Yu_SI}:
\begin{equation}\label{so}
u_{he, +\frac{3}{2}}=-\frac{1}{\sqrt{2}}(X+\mathrm i Y)\uparrow,\quad u_{he, -\frac{3}{2}}=-\frac{1}{\sqrt{2}}(X-\mathrm i Y)\downarrow,
\end{equation}
where $X$, $Y$ and $Z$ are functions of p-orbitals forming the conduction bands of perovskites.

The interband momentum matrix element $P_{cv}$ is defined as
\begin{equation}\label{He}
P_{cv}=\langle S|\hat{p}_x|X\rangle=\langle S|\hat{p}_y|Y\rangle=\langle S|\hat{p}_z|Z\rangle,
\end{equation}
where $\hat p_{x,y,z}$ are components of the momentum operator.
We are considering $\sigma^\pm$ polarized light, so we are interested in excitons with dipole moments along $x$ and $y$. To calculate the matrix element of interaction with light one has to consider functions \eqref{so} or \eqref{He}. The Kane energy $E_p$ is connected with $P_{cv}$ as
\begin{equation}
E_p=\frac{2 P_{cv}^2}{m_0}.
\end{equation}

From one point the exciton longitudinal-transverse splitting is the  consequence of the long-range part of the exchange interaction~\cite{Denisov1973_SI,Andreani1988_SI}, from the other it is connected to the exciton oscillator strength, describing the exciton contribution to the polarization induced by incoming light. Calculating the excitonic contribution to polarization analogously to Ref. [\onlinecite{Ivchenko_book_2007_SI}] we arrive to 
\begin{equation}\label{ELT3}
\hbar\omega_{LT}=\frac{4e^2\hbar^2}{3m_0 E_T^2}\frac{E_p}{\varepsilon_b a_B^3}, 
\end{equation}
Note, that there is additional factor ``2/3'' as compared with Ref. [\onlinecite{Ivchenko_book_2007_SI}]  due to the perovskite band structure.

In the quasicubical approximation the effective masses of electrons and hole can be calculated as\cite{kirstein2021nc_SI}:
\begin{equation}\label{masses}
\frac{1}{m_e}=\frac{1}{m_0}+\frac{E_p}{3m_0E_g},\quad -\frac{1}{m_h}=\frac{1}{m_0}-\frac{E_p(3E_g+\Delta)}{3m_0E_g(E_g+\Delta)},
\end{equation}
where $\Delta$ is the spin orbit splitting of the conduction band in the quasicubical approximation. One can estimate the Kane energy $E_p$ both from Eq.~\eqref{ELT3} and from Eq.~\eqref{masses}. Respective numbers are given in main text.

\end{widetext}

\end{document}